\documentclass[a4paper,12pt]{article}

\usepackage[top=2.6cm,bottom=1.8cm,inner=2.75cm,outer=2.75cm,twoside]{geometry}

\usepackage[utf8]{inputenc}
\usepackage{fontenc}
\usepackage[english]{babel}
\usepackage[dvips]{graphicx}
\usepackage{amsmath,amssymb,amsfonts}
\usepackage[usenames]{color}
\usepackage{indentfirst}
\usepackage{makeidx}
\usepackage{nopageno}
\usepackage{enumerate}
\usepackage{mathptmx}

\usepackage[small,compact]{titlesec}

\usepackage[squaren]{SIunits}

\usepackage{titling}
\predate{}%
\postdate{}%

\usepackage{authblk}
\usepackage[square,comma,sort&compress,numbers]{natbib}

\usepackage{textcomp}

\date{}

\title{Russian interbank networks: main characteristics and stability with respect to contagion}
\author[1]{A.V. Leonidov}
\author[2,*]{E.L. Rumyantsev}

\affil[1]{P.N. Lebedev Physical Institute, MIPT and ITEP, Moscow}
\affil[2]{MIPT}
\affil[*]{settlemen@mail.ru}

\begin{document}

\maketitle
\begin{abstract}
Systemic risks characterizing  the Russian overnight interbank market from the network point of view are analyzed.
\end{abstract}
\section*{Introduction}
The continuing financial crisis has focused particular attention to systemic risks related to interbank network. The corresponding literature includes papers analyzing real interbank networks \cite{ElsingerLeherSummer,UpperWorms,SantosCont}, theoretical discussions and modelling \cite{Haldane,GaiKapadia} and discussion of prudential measures \cite{BIS}. At the conceptual level the research in this area is based on a theory of complex networks, see e.g. \cite{AlbertBarabasi,DorogovtsevMendes,BocalettiLatoraMoreno}.  

The main goal of the present study is to examine the structure of the Russian interbank network and the corresponding systemic risks related to possible default of one of the banks and the volume of contagion triggered by this event. 
 
In our analysis we use the data on overnight interbank transactions of 767 banks from August 1 2011 till November 3 2011.  The choice was made in such a way that each bank had at least one transaction within the considered time period. Only transactions corresponding to borrowing (lending) money without any collateral were taken into account.

\section*{Network structure}
Let us turn to a more detailed description of the global properties of the Russian interbank networks. As has been already mentioned, the database includes all 767 banks having at least one transaction within the considered period of 69 days. 

The network of interbank interactions is defined as follows. The nodes of a network stand for banks. A (directed) link between two nodes describes an interbank interaction involving two parties, a borrower and a lender. In what follows we use a standard definition where the link is directed from a borrower to a lender. In network terms lending money to a counterparty creates an outgoing link and borrowing money - an incoming one. In addition, each link is characterized by the amount of money borrowed (lent). An interbank network is thus fully characterized by a directed weighted graph $G^{W}=(N,W)$, where $N$ is the number of the nodes and $W=\{w_{ij}\}$ is an $N\times N$ matrix of interbank exposures where $w_{ij}>0$ is a total obligation of the bank $i$ to the bank $j$.  

Let first analyze the gross geometrical features of the interbank network under consideration.  

The simplest characteristics of a network is a probability $p$ of having a link which, for a network with $N$ vertices and $K$ links can be estimated  as
\begin{equation}
p \; = \; \frac{2K}{N(N-1)}
\end{equation}
For the Russian interbank network under consideration the average value of $p$ is $\langle p \rangle \sim 0.0037$.

Another important characteristics of a graph is a clustering coefficient $C$\footnote{For simplicity in computing $C$ we treat the graph as an undirected one.} which can conveniently be defined as a ratio of a number of actually existing links between the $z$ nearest neighbors of a 
vertex and their total possible number $z(z-1)/2$. It is clear that for a totally random graph one has $C=p$. For the network under consideration the 
averaged clustering coefficients for incoming and outgoing clusters $C^{In}=0.035$ and $C^{Out}=0.012$ respectively showing a significant amount of clustering.  

Let us now turn to the global characteristics of the trading pattern corresponding to a characteristic interbank network. The simplest characteristics of an overall activity is the mean number of banks that are active or, equivalently, the mean number of banks that are passive on a given day. The corresponding values are 470 and 297 respectively, so that on a typical day we have a network of 470 active banks. These latter can be active in a different fashion. At two opposite poles are pure lenders, i.e. vertices with incoming links only - on average, 299 vertices per day and pure borrowers, i.e. vertices with outgoing links only - on average, 92 vertices. The remaining 79 vertices serve as connectors between borrowers and lenders, i.e. borrow and lend at the same time. A more detailed description of an average daily "in-out" pattern is presented in the Table \ref{tab:decomposition}.
\begin{table}[h]
\caption{Interbank market structure}
\label{tab:decomposition}
\begin{center}
\begin{tabular}{|c|c|c|c|c|}
\hline
Condition & $k=0$ & $k>0$ & $k>2$ & $k>10$\\ \hline In & $389$ $(14)$ & $378$ $(14)$ & $126$ $(9)$ & $14$ $(3)$\\ \hline Out & $596$ $(8)$ & $171$ $(8)$ & $82$ $(5)$ & $29$ $(3)$\\ \hline Only In & $297$ $(15)$ & $299$ $(14)$ & $82$ $(8)$ & $2$ $(1)$\\ \hline Only Out & $297$ $(15)$ & $92$ $(8)$ & $29$ $(4)$ & $5$ $(2)$\\ \hline 
\end{tabular}
\end{center}
\end{table}
The columns in Table \ref{tab:decomposition} correspond to an average daily number of vertices of given type (In, Out, etc.) satisfying some certain condition. In parentheses we show the corresponding standard deviations.

As has been already mentioned, the systemic risk associated with an interbank network refers to propagation of defaults triggered by the default of one or several banks (vertices) and propagating along outgoing links. In this context the properties of the Russian interbank market characterized by the Table  \ref{tab:decomposition} lead us to the following observations:
\begin{enumerate}
\item The number of pure lenders is almost twice as large as that of pure borrowers. This feature creates specific systemic risks  because a default of any borrower may lead to defaults of several lenders.
\item Of special interest are those 29 banks which are characterized by large ($k_{\rm out}>10$) values of their out-degree. These banks are clearly especially important sources of systemic risk. Let us note that these banks accumulate $63\%$ of the total systemic debt.
\item Let us also point at those $14$ banks that have more than $10$ incoming links. From the network perspective these banks play a role of hubs absorbing potential shocks due to their loan's diversification. These banks control $37\%$  of the total loan.
\end{enumerate} 

\begin{figure}[ht]
\centering
\includegraphics[angle=-90,width=0.99\textwidth]{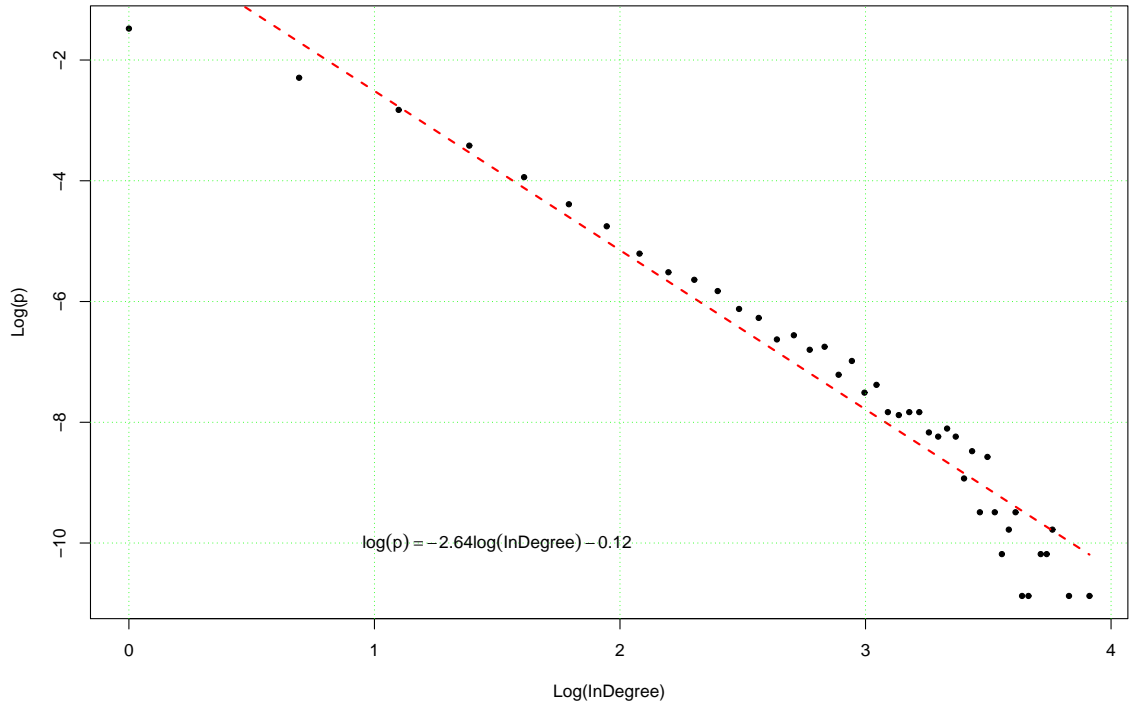}
\caption{\small Marginal in- degree distribution.}
\label{ris:MargInDistr}
\end{figure}

\begin{figure}[ht]
\centering
\includegraphics[angle=-90,width=0.99\textwidth]{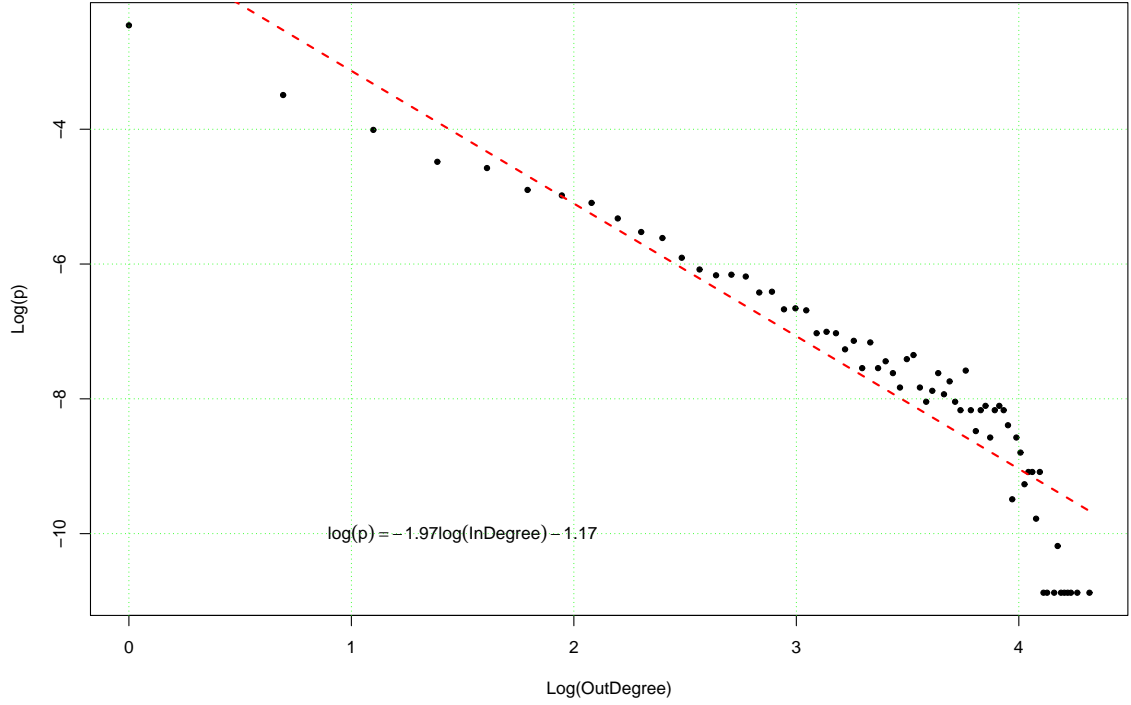}
\caption{\small Marginal out- degree distributions.}
\label{ris:MargOutDistr}
\end{figure}

An important generic feature of a directed network are the probability distributions $P(k^{\rm \; in})$ and $P(k^{\rm out})$ for the number of incoming and outgoing links for a vertex. The majority of networks discussed in the literature, see e.g. \cite{AlbertBarabasi,DorogovtsevMendes,BocalettiLatoraMoreno}, are the so-called scale-free ones, i.e. have powerlike tails $P(k)\sim {\rm const} / k^{\gamma}$. The corresponding marginal in- and out- degree distributions for the Russian interbank market are shown in Figs. \ref{ris:MargInDistr} and \ref{ris:MargOutDistr} respectively. We  see that the network is scale-free for both distributions with $\gamma^{\rm \; in}=1.92$ and $\gamma^{\rm out}=2.64$.

Propagation of contagion in a network is crucially dependent on its connectivity. The simplest corresponding characteristics is an average number of in- and out- links $z_2^{\rm \; in}$ and $z_2^{\rm \; out}$ of the nearest neighbors of a vertex. The bigger are  $z_2^{\rm \; in}$ and $z_2^{\rm \; out}$ in comparison with the mean number of in- and out- links for a vertex $z_1 = z_1^{\rm \; in} = z_1^{\rm \; out}$, the easier is contagion propagation along the corresponding cluster. For the Russian interbank network we have $z_1=1.41$, $z_2^{\rm \; in}=9$ and $z_2^{\rm \; out}=28$. This the condition $z_2/z_1>2$ for an existence of a giant component holds both for in- and out- clusters. 

\begin{figure}[ht]
\begin{minipage}[h]{0.49\linewidth}
\center{\includegraphics[height=0.35\textheight,width=0.99\textwidth]{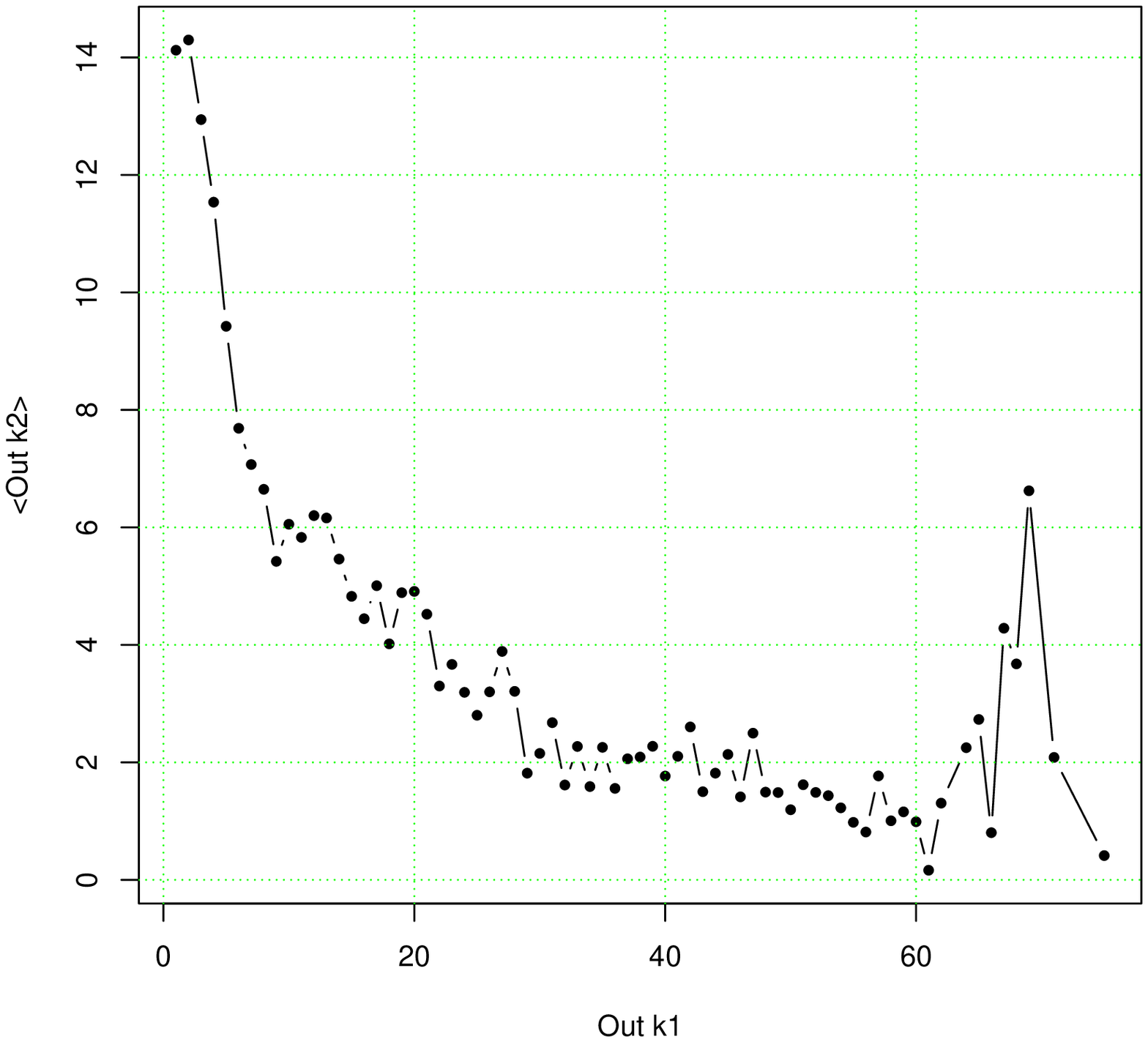} \\ a}
\end{minipage}
\hfill
\begin{minipage}[h]{0.49\linewidth}
\center{\includegraphics[height=0.35\textheight,width=0.99\textwidth]{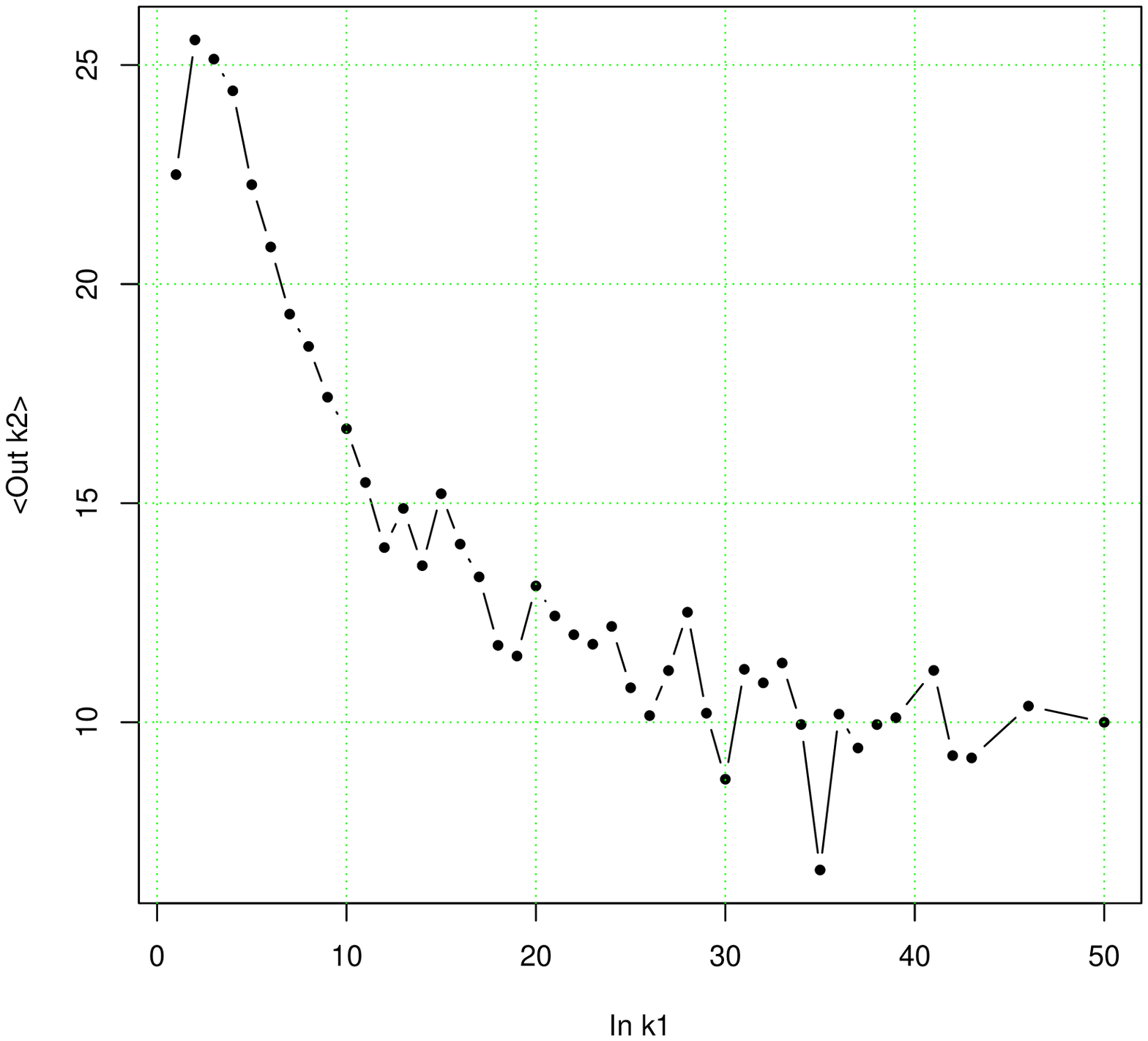} \\ b}
\end{minipage}
\caption{\small Out-in (a) and out-out (b) degree correlations}
\label{ris:DegreeCorrelation}
\end{figure}

A more detailed information on the network connectivity is given by a conditional probability distribution $P \left( k_2^{\rm \; in},k_2^{\rm \; out} \vert k_1^{\rm \; in},k_1^{\rm \; out} \right)$ characterizing probability for a nearest neighbor of a vertex with $k_1^{\rm \; in}$ incoming and  $k_1^{\rm \; out}$ outgoing to have $k_2^{\rm \; in}$ and $k_2^{\rm \; out}$ incoming and outgoing links respectively. In Fig. \ref{ris:DegreeCorrelation} we show two lowest moments of its marginal distributions, namely $ \langle k_2^{\rm \; out} \rangle (k_1^{\rm \; out})$, Fig. \ref{ris:DegreeCorrelation} (a), and $ \langle k_2^{\rm \; in} \rangle (k_1^{\rm \; out})$, Fig. \ref{ris:DegreeCorrelation} (b). Both plots show pronounced assortiativity at small  $k_1^{\rm \; out}$ for both $k_2^{\rm \; in}$ and $k_2^{\rm \; out}$.

\section*{Contagion effect}

\begin{figure}[h]
   \begin{center}
     \includegraphics[width=0.7\textwidth]{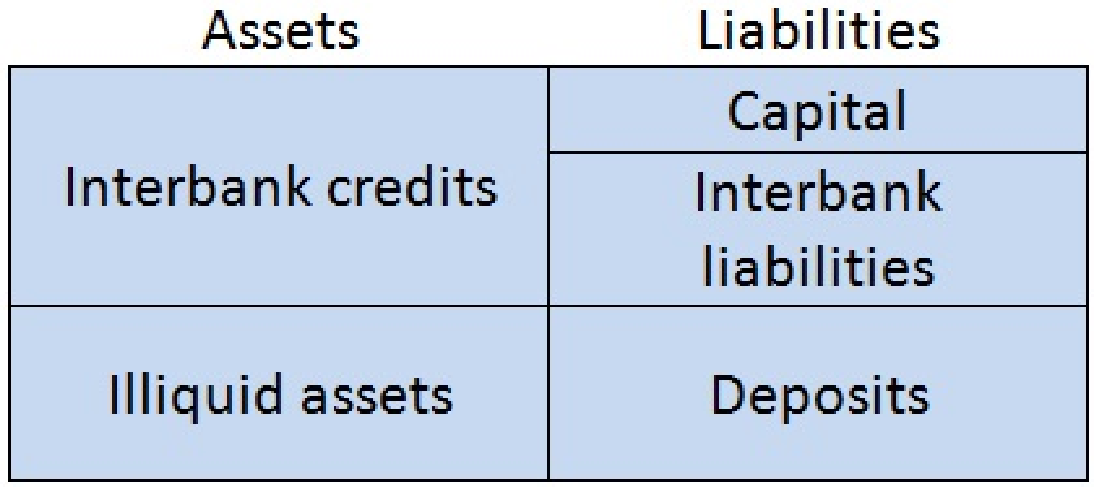}
   \end{center}
   \caption{\small Simplified balance sheet}
   \label{fig:BalanceSheet}
 \end{figure}

Let us start with formulating the model of default contagion spreading in a bank network we are using in the present study. The original source of risk are banks (vertices) that have loans and default on their payment. The banks immediately affected by such a default are the nearest neighbors of this vertex reachable via outgoing links attached to it. If one of the nearest neighbors also defaults, the process can spread further. A probability of infection depends on the number of incoming and outgoing links of a vertex. In the present study we use a simple stylized model of bank balance sheets from \cite{GaiKapadia}. However, at difference with the analysis of \cite{GaiKapadia} and similarly to \cite{SantosCont}, we are working with the real day-by-day topologies of the interbank market. In this model a representative bank has a simple balance sheet structure with interbank and illiquid assets on the assets side and capital, deposits and interbank obligations on the liabilities side shown in Fig. \ref{fig:BalanceSheet}.
The corresponding condition for the bank $i$ to be solvent is 
\begin{equation}
   (1-\phi)A_{i}^{IB}+qA_{i}^{M}-L_{i}^{IB}-D_i > 0,
   \label{eq:SolvencyCondition}
\end{equation}
where $A_{i}^{IB}$ denotes interbank assets of bank $i$, $A_{i}^{M}$ -- its illiquid assets, $L_{i}^{IB}$ -- its interbank liabilities,  $\phi$ is a fraction of banks having obligations with respect to bank $i$ that have defaulted, and $q$ is the discount for fire-selling illiquid assets. It is assumed  that interbank claims and liabilities of a particular bank are uniformly distributed across its borrowers and creditors so that $\phi=\frac{1}{j_i}$ where $j_i$ is the number of borrowers. This assumption highlights an importance of the incoming degree for each bank as reflecting its risk diversification. Banks with a high value of the incoming degree have lower probability to go bankrupt due to contagion effect. At the same time banks with a high value of  the outgoing degree can be the sources of contagion. For simplicity neglect the fire discount, i.e. assume  $q=1$ in Eq. \ref{eq:SolvencyCondition}. Therefore one can rewrite the solvency condition \ref{eq:SolvencyCondition} as follows:
\begin{equation}
   \frac{K_i}{A_{i}^{IB}}>\frac{1}{j_i},
   \label{eq:DefaultCondition}
\end{equation}
where the capital buffer $K_i$ is defined as
\begin{equation}
K_i \; = \; A_{i}^{IB}+qA_{i}^{M}-L_{i}^{IB}-D_i
\end{equation}

Our main goal will be to study the impact of the capital buffer size on the number of banks which default due to contagion. The procedure we use is as follows:
 \begin{enumerate}
   \item We set the values of relevant parameters. We assume that $A_{i}^{IB}$ makes 20\% of the balance sheet and study a range of capital buffer values from 4 to 10 \% of the balance sheet. 
   \item Taking real structure of the overnight interbank market we default each bank and determine the size of the default cluster by checking, using Eq. \ref{eq:DefaultCondition}, whether some of its nearest neighbors that can be reached from the defaulted vertex via outgoing links are infected, etc. For each initial bank $i$ a default cluster for  is the number of banks defaulted due to the default of $i$ as a result of contagion process. 
   \item Finally, for each value of the capital buffer we calculate an average over default cluster.
 \end{enumerate}

\begin{figure}[ht]
 \begin{minipage}[h]{0.49\linewidth}
   \begin{center}
     \includegraphics[width=0.99\textwidth]{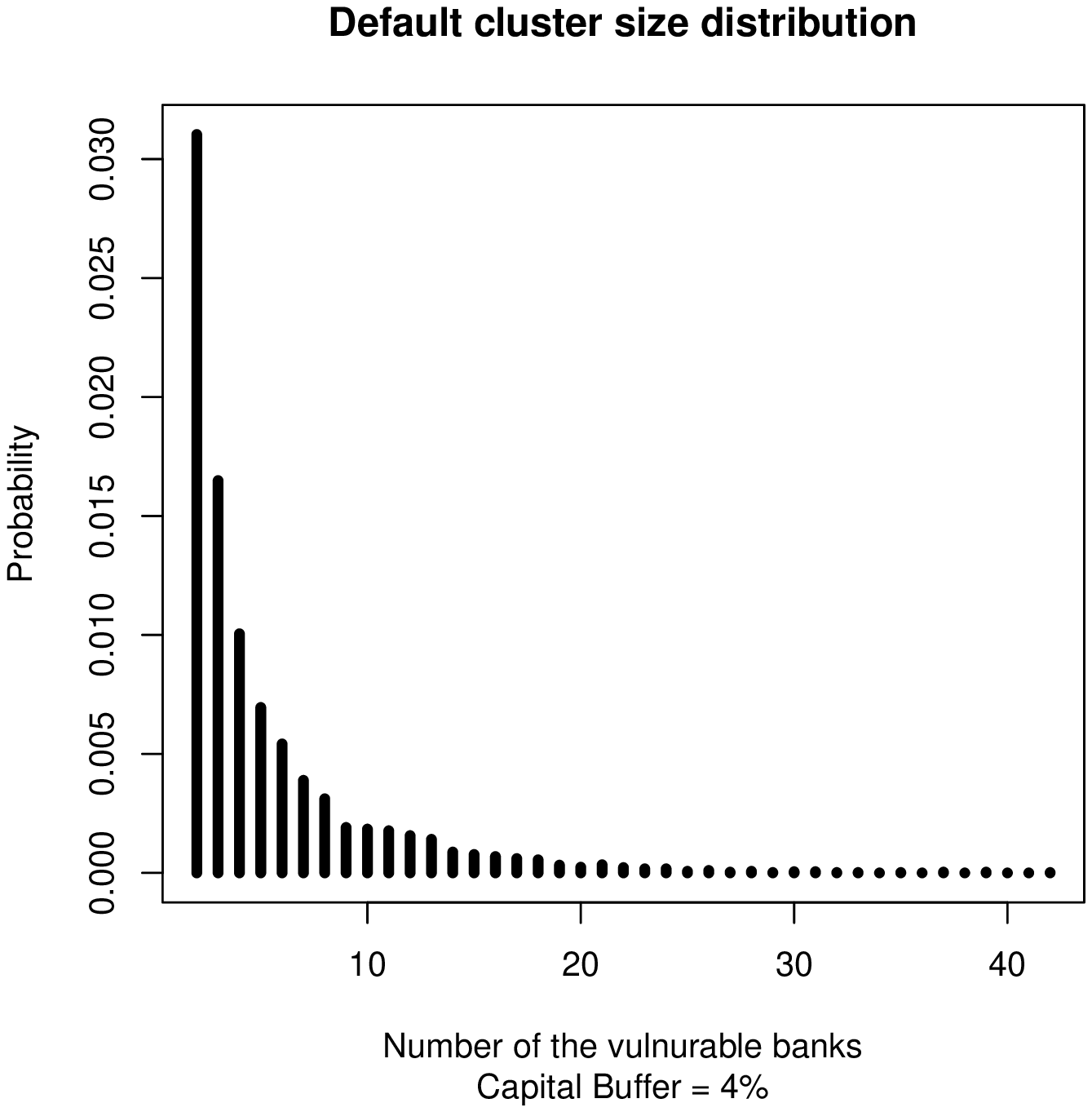}
   \end{center}
   \caption{\small Probability distribution for cluster size, capital buffer 4\%.}
   \label{fig:ProbDistrib}
 \end{minipage}
\hfill
 \begin{minipage}[h]{0.49\linewidth}
   \begin{center}
     \includegraphics[width=0.99\textwidth]{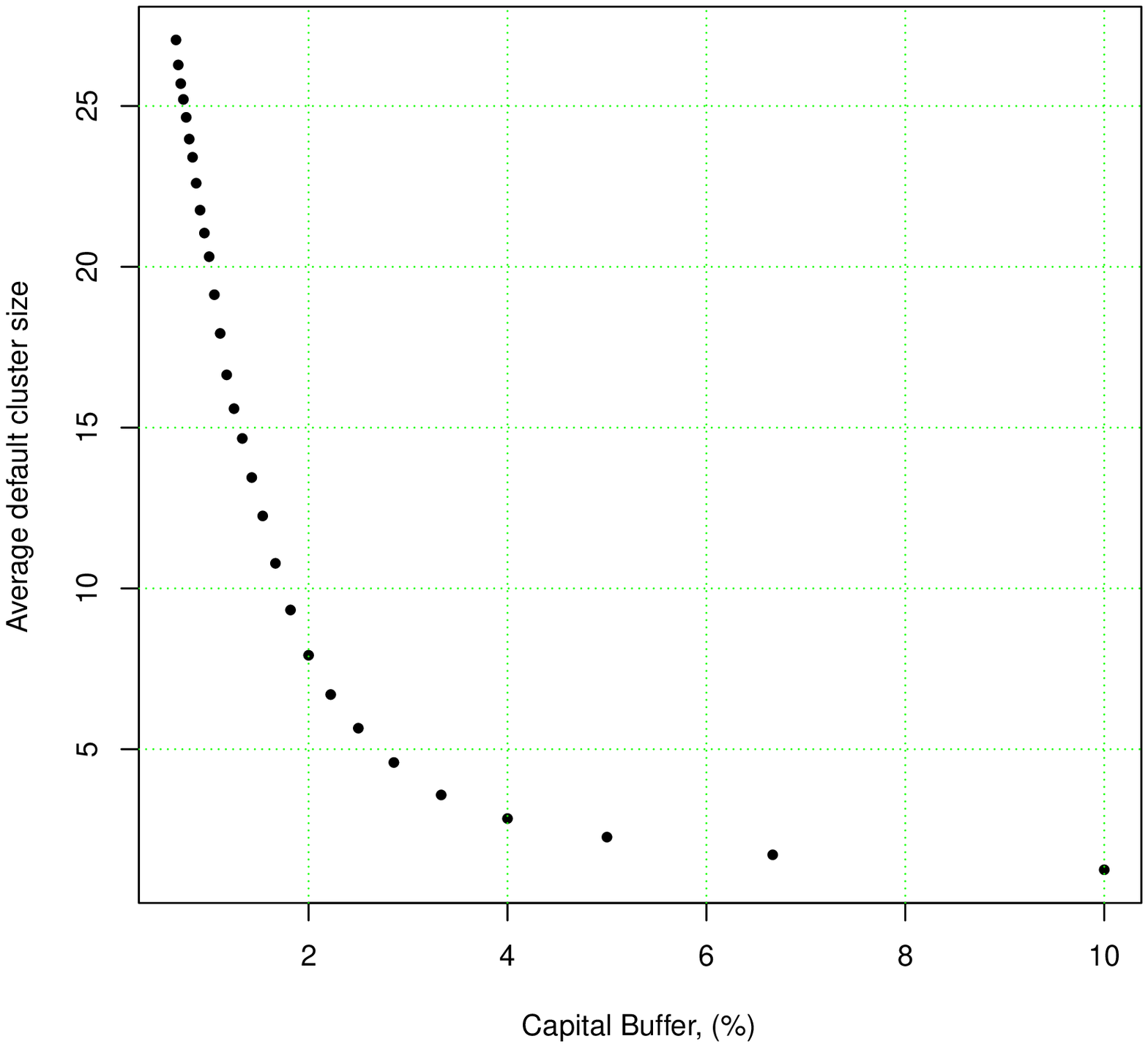}
   \end{center}
   \caption{\small Average size of default cluster as a function of a capital buffer.}
   \label{fig:DefClusDistrib}
   \end{minipage}
 \end{figure}

The results of this simulation are presented in Figs. \ref{fig:ProbDistrib} and \ref{fig:DefClusDistrib} in which probability dsitribution of default cluster sizes (Fig. \ref{fig:ProbDistrib}) and a dependence of the average default cluster on the value of capital buffer (Fig. \ref{fig:DefClusDistrib}) are shown. 
From the distribution in  Fig. \ref{fig:ProbDistrib} we can make quantitative statements on the significance of systemic network-related risks. For example, for the capital buffer size of $4\%$ of total balance sheet there is a $1\%$ probability for more than $8$ banks go bankrupt. The main conclusion that can be drawn from Fig. \ref{fig:DefClusDistrib} is that the average default cluster size is rapidly decaying with growing capital buffer. 
\section*{Conclusion}
In this study we have analyzed some systemic network-related properties of the Russian overnight interbank market. A detailed analysis will be published in \cite{RL12}.  
% insert bibliography if needed
%\bibliography{ICENET-2012_biblio_template}

\end{document}